\documentclass[aps,twocolumn,showpacs]{revtex4}
\usepackage{graphicx}%
\begin{document}
\title{DNA denaturation bubbles at criticality }
%
%
%
\author{Nikos Theodorakopoulos$^{1,2}$ } 
\affiliation{
$^{1}$Theoretical and Physical Chemistry Institute, National Hellenic Research Foundation,
Vasileos Constantinou 48, 116 35 Athens, Greece \\
$^{2}$Fachbereich Physik, Universit\"at Konstanz, 78457 Konstanz, Germany\\
}
\date{\today}
\begin{abstract}
The equilibrium statistical properties of DNA denaturation bubbles are examined in detail within the framework of the Peyrard-Bishop-Dauxois model. Bubble formation in homogeneous DNA is found to depend crucially on the presence of nonlinear base-stacking interactions. Small bubbles extending over less than 10 base pairs are associated with much larger free energies of formation per site than larger bubbles. As the critical temperature is approached, the free energy associated with further bubble growth becomes vanishingly small. An analysis of average displacement profiles of bubbles of varying sizes at different temperatures reveals almost identical scaled shapes in the absence of nonlinear stacking; nonlinear stacking leads to distinct scaled shapes of large and small bubbles. 
\end{abstract}
\maketitle
\section{Introduction}
The nonlinear dynamics and statistical physics of DNA denaturation have been widely investigated \cite{PeyrRev}. 
Recent work has attempted to extend mesoscopic scale modeling in order to describe with sufficient accuracy how sequence details determine the statistical and dynamical properties of local fluctuations. Such local fluctuations, known as \lq\lq denaturation bubbles\rq\rq\- are believed to be instrumental in the {\em initiation} of the transcription process at physiological temperatures. The possibility of spontaneous, sequence-specific formation of a mid-size bubble has been the subject of considerable research interest - and some debate - \cite{Choi04, Kalo04, vanErp, BishComm, PeyrComm}. A related - but nonetheless distinct - question which might be relevant to the {\em process} of transcription concerns the growth of a bubble to much larger sizes. Since this is by definition - at least in an asymptotic sense - a scale-free phenomenon, it is best addressed at the level of the underlying phase transition; moreover, at least its salient features should be evident within the context of the homogeneous (polynucleotide) chain.

%
%
It should be recalled that nonlinear lattice dynamics based DNA modeling of the Peyrard-Bishop-Dauxois (PBD \cite{PBD}) type predicts either an (effectively) first-order or a (strict) second-order phase transition, depending on whether non-linear base-stacking effects are taken into account or not \cite{TDP}. In the case of second-order transition, it has been determined that domain walls (DW) become entropically stable at the critical temperature \cite{JSP2001, Physica2006}; recent numerical evidence from Monte-Carlo simulations \cite{SungPRE04} suggests that the average bubble size also becomes critical. The first-order transition case is slightly more complicated. Entropic effects are not sufficient to enable spontaneous DW formation at the critical temperature. This appears to rule out DWs as agents of thermal denaturation. Bubbles are natural - and in fact have always been - prime suspects for this role. Very recent work \cite{Kalo2007} has demonstrated that large bubbles may form in this case as well, and that their probability distribution cannot be described by a simple exponential. Detailed data in the vicinity of the critical temperature are not available\cite{foot}. It is however known from previous work \cite{Hwa,TDP} that most of the physics of nonlinear base-stacking is generated by an effective thermal barrier which modifies the on-site Morse-like potential. The point at which the effects of the thermal barrier become important defines a natural crossover between different types of behavior.  As this note will show in some detail, such a crossover is also present in bubble statistics. Bubbles which extend over a few sites are entirely non-critical; the onset of criticality is reflected only in the statistics of large bubbles; the asymptotic properties of the latter are such that 
the free-energy barrier toward bubble growth is lowered - approaching zero at the transition temperature, whereas the average bubble size always remains finite. It will be further shown that this dichotomy between large and small bubbles is not restricted to the statistics; reduced shapes of large and small-size bubbles - which scale uniformly in the absence of nonlinear stacking interactions - reveal distinct differences when nonlinear stacking is included.

The paper is organized as follows: Section II includes model definitions, notation and general properties of bubbles. Section III presents numerical results on bubble statistics based on direct matrix multiplication. Section IV formulates an alternative procedure based on an associated eigenvalue problem which provides emphasis on asymptotic properties. Section V discusses bubbles shapes. The final section includes a brief summary and discussion of some key points. 

\section{Definitions}
\subsection{Model}
The PBD model assumes a potential energy of the form
\begin{equation}
	H_P = \sum_{j=1}^N \left\{ W(y_{j-1},y_j)  + V(y_j) \right\}
\end{equation}
where $y_j$ is a transverse coordinate representing the separation of the two bases at the $j$th site, 
\begin{equation}
	W(y,y') = \frac{1}{2R}\left[ 1 + \rho e^{-b(y+y')}  \right](y-y')^2
\end{equation}
is an anharmonic elastic term which models the nonlinear base-stacking interaction and 
$V(y) = (1-e^{-y})^2$
is an on-site Morse potential describing the combined effects of hydrogen-bonding, stacking and solvent. I will use the dimensionless parameter values $R=10.1$, $b=0.08$ and, unless otherwise stated, $\rho=1$. Furthermore, I will assume that the system is subjected to periodic boundary conditions. Thermodynamics is governed by the properties of the transfer integral (TI) equation 
\begin{equation}
	\int_{-\infty}^{\infty} dy'K(y,y') \phi_{\nu}(y') = \Lambda_\nu \phi_{\nu}(y)
\end{equation}
with $K(y,y')=e^{-[W(y,y')+V(y)/2+V(y')/2]/T}$ and $T$ the dimensionless temperature. In particular, details of a possible phase transition depend on the type of singularity (if any) which the spectral gap $\Delta \epsilon = - T \ln ( \Lambda_1 / \Lambda_0)$ might exhibit near a critical temperature $T_c$. Since the spectral gap is equal to the singular part of the thermodynamic free energy per site \cite{JSP2001}, a linearly vanishing gap as $T\to T_c^-$ corresponds to a first order transition, a quadratically vanishing gap to a second order transition etc.
\subsection{Bubbles}
The $n$th base pair is assumed to be unbound if $y_n > y_c$; it is in a bound state if $y \leq y_c$; I choose $y_c=\ln 2$, the inflexion point of the Morse potential. The choice is of course somewhat arbitrary, but it should not influence fundamental asymptotic results. A bubble of length $n$ is a sequence of $n$ successive unbound sites preceded and followed, respectively, by a single bound site. It is present in the infinite system (assumed subjected to periodic boundary conditions) with a probability
\begin{eqnarray}
\nonumber
	P_n & = & \lim_{N\to \infty} \frac{1}{Z_N}\int_{-\infty}^{\infty} dy_1 \cdots  dy_{r-1} 
	\int_{-\infty}^{y_c}dy_r \int_{y_c}^{\infty}dy_{r+1}
	\\
\nonumber
	&& 	  \cdots  dy_{r+n} \int_{-\infty}^{y_c} dy_{r+n+1} \int_{-\infty}^{\infty} dy_{r+n+2} \cdots dy_{N} \\
	&&  	  K(y_1,y_2)\cdots K(y_N,y_1) \\
& = & 	\int_{-\infty}^{y_c}dy_r \phi_0^*(y_r)
	 \int_{y_c}^{\infty}dy_{r+1} \cdots  dy_{r+n} \int_{-\infty}^{y_c} dy_{r+n+1}  \nonumber \\
	&& \phi_0(y_{r+n+1}) {\hat K}(y_r,y_{r+1})\cdots {\hat K}(y_{r+n},y_{r+n+1})	
	\label{eq:bubbleprob}
	 \end{eqnarray}
where $Z_N$ is the full configurational partition function, dominated by the highest eigenvalue $\Lambda_0$,
$\phi_0$ denotes the TI eigenstate corresponding to $\Lambda_0$, and ${\hat K} = K/\Lambda_0$. 

\subsection{Sum rules}
By definition, the sum of all $P_n$'s expresses the probability that the site which precedes the bubble has a  bound base pair, i.e.
\begin{equation}
	\sum_{n=0}^{\infty} P_n = p = \int_{-\infty}^{y_c} dy \>  	|\phi_0(y)|^2 \quad.
	\label{eq:boundsites}
\end{equation}
Moreover, the sum 
\begin{equation}
	\sum_{n=1}^{\infty}n P_n = 1 - p 
\label{eq:unboundsites}
\end{equation}
expresses the fraction of sites with unbound base pairs. As a consequence, the average bubble size (including the correct weighting factor for bubbles of zero length) is 
\begin{equation}
	\xi_b = \frac{1-p}{p}  \quad.
\end{equation}
Some general conclusions can already be drawn at this level. For a second-order transition, where $p \propto T_c-T$ near $T_c$ \cite{TDP}, $\xi_b \propto (T_c-T)^{-1}$. For a first-order transition, where $p$ approaches a constant as $T \to T_c^{-}$, $\xi_b$ remains finite. Note that  the average bubble length is in both cases much smaller that the correlation length $ \xi = T/\Delta \epsilon$ which respectively diverges quadratically or linearly. \par
\begin{figure}[h]
\vskip -.3truecm
\resizebox{0.482\textwidth}{!}
{\includegraphics{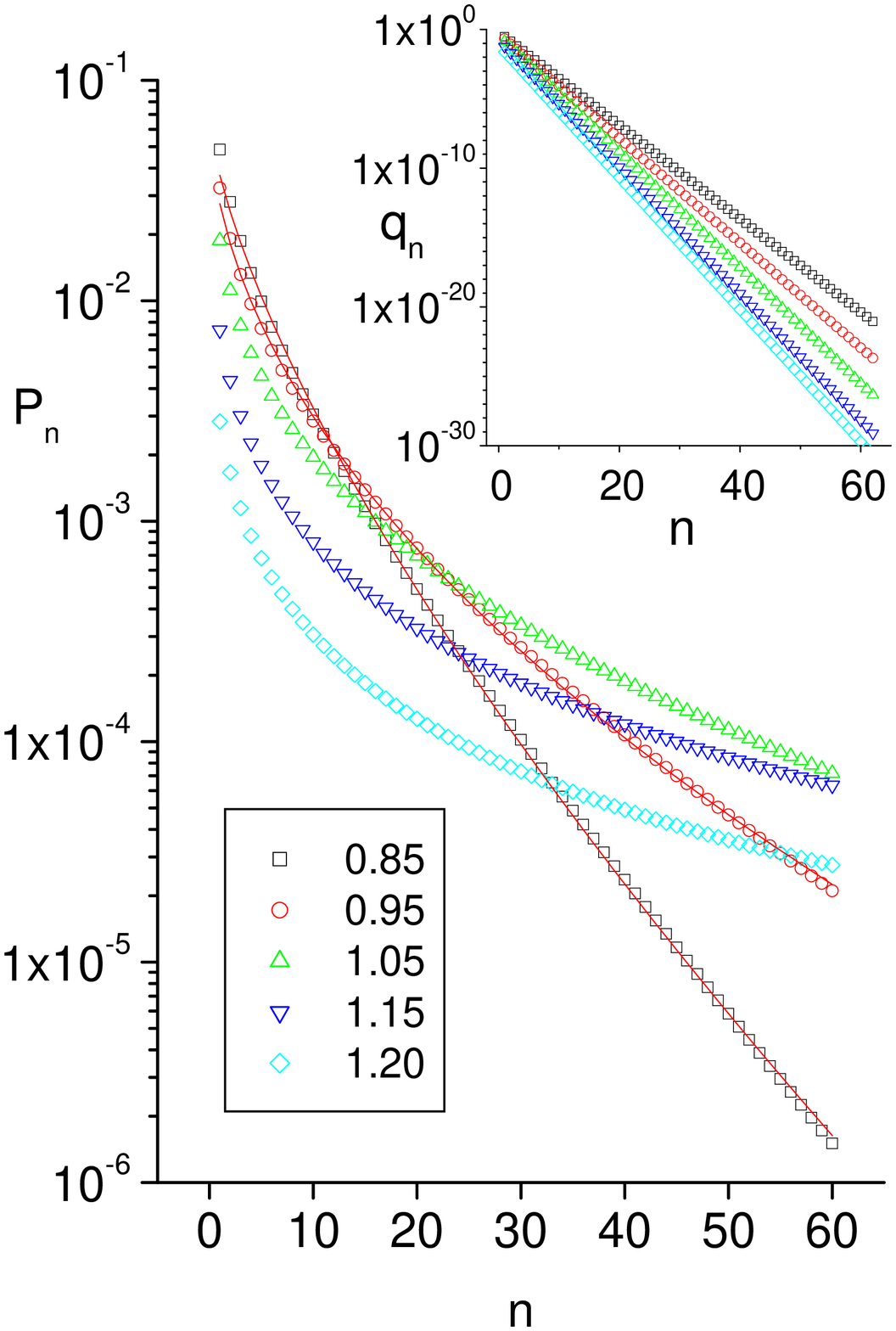}	
\includegraphics{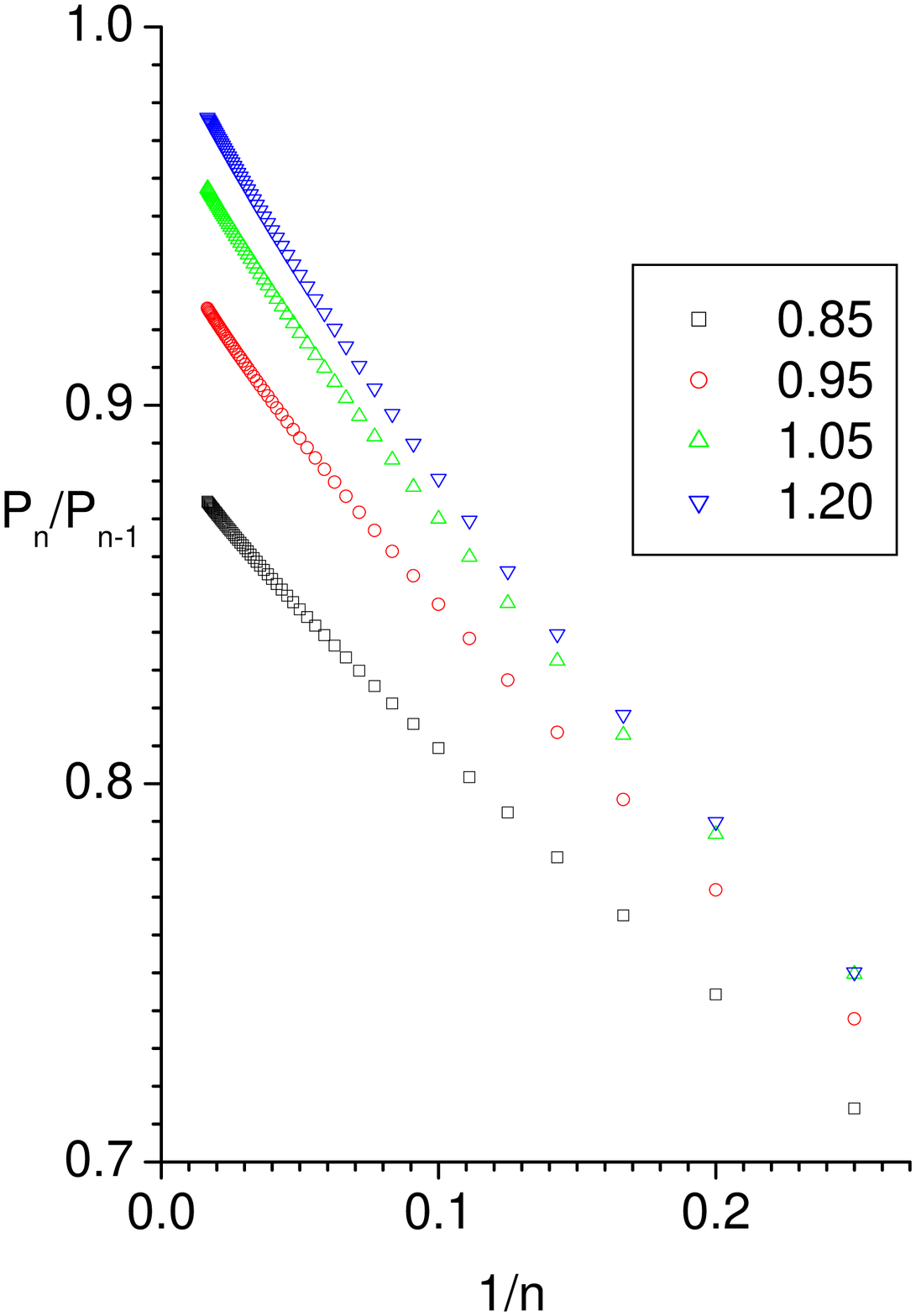}}	
\vskip -.5truecm
\caption{ 
{\small (Color online) Bubble statistics in the case of vanishing nonlinear base stacking, $\rho =0$. {\em Left panel}: Probability of a bubble extending to $n$ sites, for a variety of temperatures. Note that the decay becomes weaker as the critical temperature is approached. Also shown are, for the two lowest temperatures, fits to a stretched exponential function. Inset: For comparison, I show the probability of a cluster of $n$ succsessive sites with bound base pairs (pure exponential). 
{\em Right panel}:  Ratios of successive probabilities vs. inverse bubble size for a range of temperatures.
}}
\label{fig:PB1}
\end{figure}
\section{Results}
For bubbles of up to moderately large size, it is possible to obtain results by direct matrix multiplication of (\ref{eq:bubbleprob}). I use a grid of 2989 points and perform the successive integrations using a 10th order Bode\cite{Abramowitz} routine in the interval $(-5,205)$.
\subsection{Linear base-stacking ($\rho=0$)} 
\begin{figure}[h]
\centering
	\vskip -.3truecm
\resizebox{0.35\textwidth}{!}
{		\includegraphics{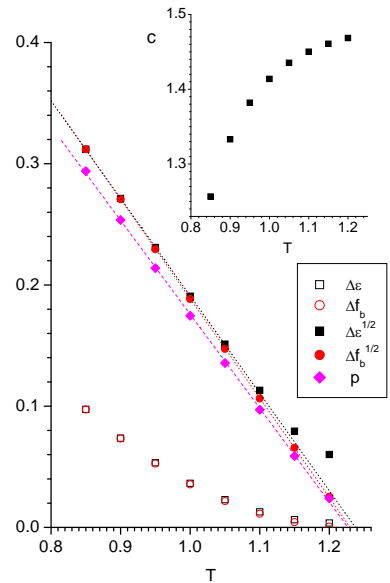}}
\caption{ 
{\small 
(Color online) Summary of critical results in the case $\rho=0$: 
as extracted from the numerical TI solution and the asymptotic behavior of the curves shown in the right panel of 
Fig. \ref{fig:PB1}: (i) spectral gap $\Delta \epsilon$ (open squares) vs. temperature, (ii) activation free energy $\Delta f_b$ associated with bubble growth (open circles); (iii) the square root  $\Delta \epsilon ^{1/2}$ (full squares) is known to depend linearly on the temperature; rounding is due to the finiteness of the matrix used; (iv) $\Delta f_b^{1/2}$ (full circles) is found to depend linearly on the temperature; remarkably, it exhibits no rounding; (v) the fraction of bound sites $p$ (diamonds), as obtained from (\ref{eq:boundsites}). The dotted lines represent linear fits to the data (cf. text for discussion). Inset: the exponent $c$ vs temperature.}
} 
	\label{fig:DfPB}
\end{figure}

Fig. \ref{fig:PB1} shows the probability of bubble occurrence (left panel) for a variety of temperatures. It is clear that that the distribution is far from exponential \cite{foot0}. For comparison I show in the inset the probability of occurrence of a bound cluster of $n$ sites, which is described by a pure exponential. Note that it is possible to fit the $P_n$ data, at least at the lowest temperatures, by stretched exponentials, i.e. 
$P_n = a \exp [ -(n/\sigma)^b ] $. However, the fits are neither perfect (they show systematic deviations at both ends) nor very instructive (the $\sigma$ obtained {\em decreases} as the temperature increases, i.e. $\sigma=2.04$ at $T=0.85$,  $\sigma=1.07$ at $T=0.95$ the necessary compensation is achieved by a substantial decrease in the stretching exponent from $b=0.70$ to $b=0.52$). A more promising approach is to separate out any implicit exponential dependence by looking at the ratios of successive probabilities. Thus if one attempts to describe deviations from exponential dependence by a power-law,  
\begin{equation}
	P_n \propto \frac{1}{n^c} e^{-n/\sigma}
	\label{eq:PnPS}
\end{equation}
a plot of the ratios 
\begin{equation}
\frac{P_n}{P_{n-1}} = e^{-1/\sigma}\left\{1 - \frac{c}{n} + \cdots \right\} \quad, \quad n\gg 1
	\label{eq:RatiosPS}
\end{equation}
vs $1/n$ should approach a definite limit as $n \to \infty$, from which it is in principle possible to read off both the activation free energy $\Delta f_b=T/\sigma$ associated with bubble growth and - by estimating the asymptotic slope - the exponent $c$
\cite{foot2}. 
The successive ratios shown in the right panel of Fig. \ref{fig:PB1} confirm this picture; results of the extrapolation are summarized in 
Fig. \ref{fig:DfPB}. They show that it is indeed consistent to represent bubble statistics by (\ref{eq:PnPS}), that $\Delta f_b \propto (T_c-T)^2$ and that the exponent $c$ varies significantly with temperature, approaching a value close to $3/2$ near $T_c$. \par
An appropriate measure of the quality of the numerical data is given by the estimated critical temperatures (cf. intersections of the dotted lines in Fig. \ref{fig:PB1} with the horizontal axis). Estimates obtained, respectively, from the fraction of bound sites $p$ and from $\Delta f_b$, are $1.227(6)$ and $1.229(1)$. They should be compared with the value $1.2276(4)$ obtained via systematic finite-size scaling analysis \cite{FSS}.
\subsection{Nonlinear base-stacking}
\begin{figure}[h]
\centering
	\vskip -.3truecm
\resizebox{0.35\textwidth}{!}
{		\includegraphics{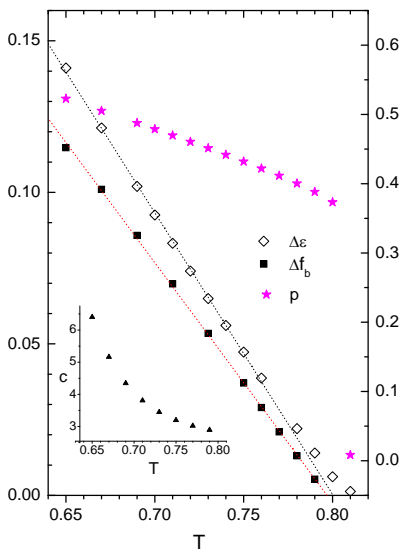}}
\caption{ 
{\small 
(Color online) Summary of critical results in the case $\rho=1$: (i) spectral gap $\Delta \epsilon$ (diamonds) and (ii)
fraction of bound sites $p$ (stars) vs. temperature, as obtained from the TI numerical solution; 
 (iii) activation free energy $\Delta f_b$ associated with bubble growth (filled squares) and (iv) exponent $c$ of the {\em asymptotic} form
 (\ref{eq:RatiosPS}) (triangles, inset), as extracted from the $n \to \infty$ asymptotics of Fig. \ref{fig:PBD}.  
  The dotted lines represent linear fits to the data (i) and (ii), both yielding, respectively, $T_c=0.80$ within numerical accuracy.
  }}
	\label{fig:Df_PBD}
\end{figure}

\begin{figure}[h]
\resizebox{0.482\textwidth}{!}
{\includegraphics{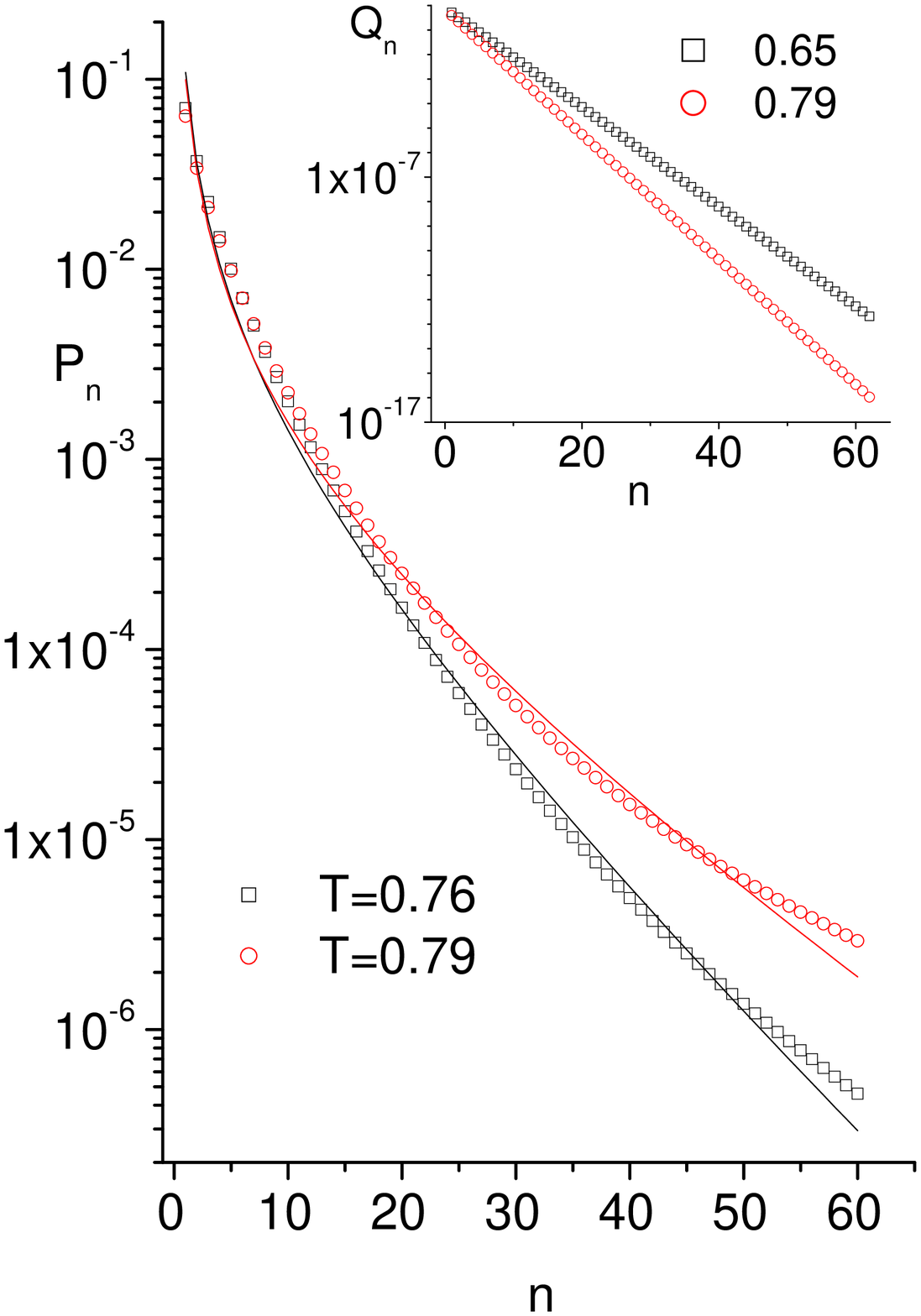}	
\includegraphics{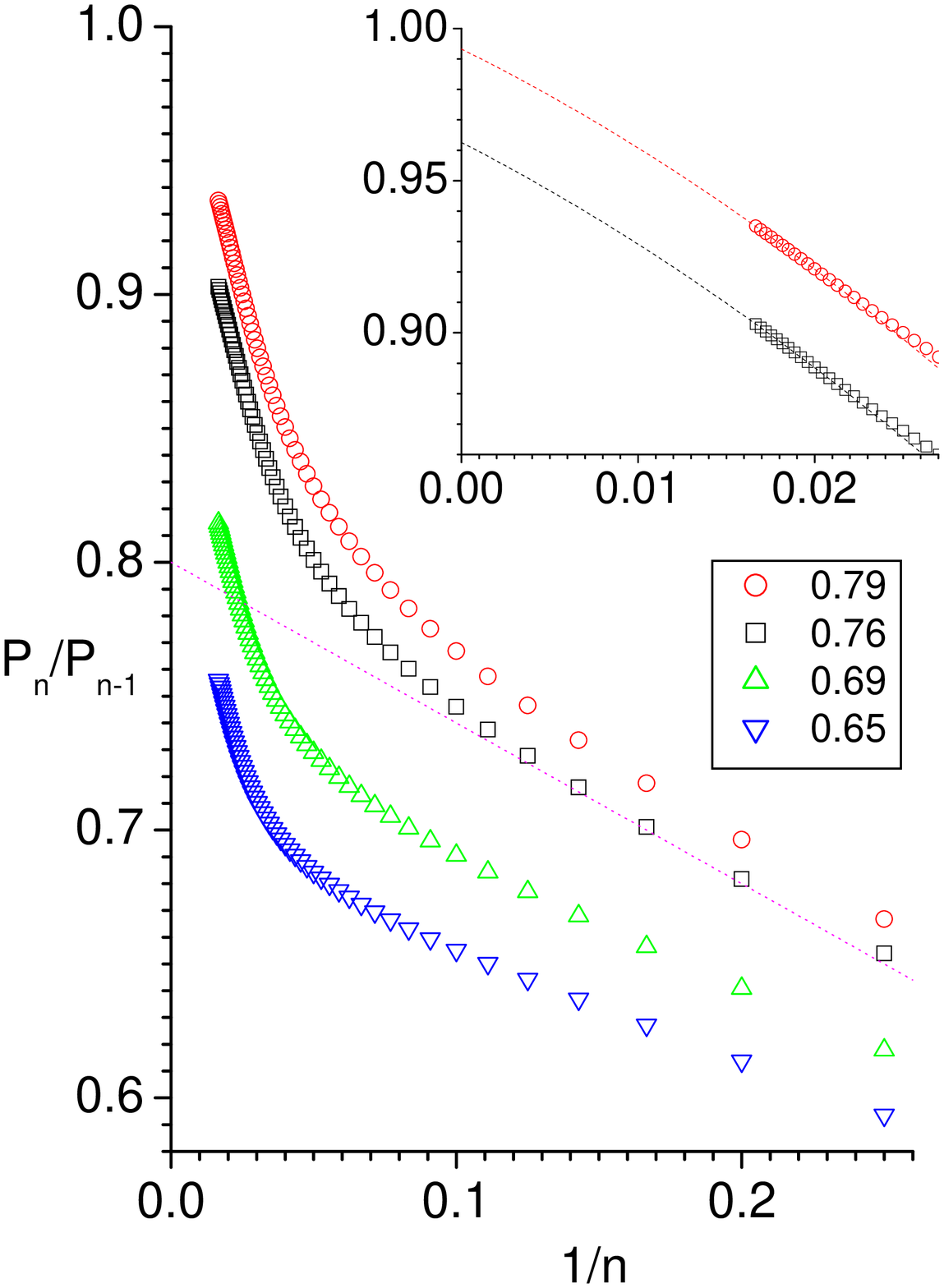}}	
\vskip -.5truecm
\caption{ {\small (Color online) Bubble statistics in the case of nonvanishing nonlinear base stacking, $\rho =1$. {\em Left panel}: Probability of a bubble extending to $n$ sites at $T=0.76$ and $T=0.79$; also shown (continuous lines) are fits to the function (\ref{eq:PnPS}), cf. text for a discussion of the fitting parameters. Inset: For comparison I show the probability of a cluster of $n$ bound sites (pure exponential). 
{\em Right panel}:  Ratios of successive probabilities vs. inverse bubble size for a range of temperatures. Inset: a zoom of the asymptotic region for the two highest temperatures; dotted lines represent quadratic extrapolations from the last 3 points. }}
\label{fig:PBD}
\end{figure}

I now proceed to the physically more relevant case of nonlinear base-stacking, $\rho=1$. Although this value underestimates the importance of nonlinear stacking interactions, it facilitates the present discussion because the relevant crossover effects occur in numerically observable regions. The important qualitative features remain unchanged. In particular, the transition is for all practical purposes a first-order one. The TI spectral gap vanishes linearly near $T_c=0.801$ and the fraction $p$ of bound sites has an apparent discontinuity at that temperature (cf. Fig \ref{fig:Df_PBD}). In order to avoid data cluttering, the left panel of Fig. \ref{fig:PBD} shows the function $P_n$ for two temperatures only, $T=0.76$ and $T=0.79$; note that the higher temperature is quite close to the critical temperature. 

It is possible to obtain rough fits to the full sets of data with the functional form (\ref{eq:PnPS}) (dotted lines); the parameter values obtained are $\sigma=8.4, 12.4$ and $c=1.42, 1.49$ - where the second value refers to the higher temperature. Such fits over the entire data range are of course of questionable value if one tries to extract asymptotic information. I have included them because the extracted parameter values provide a hint of the underlying problem: the $\sigma$'s, although much larger than unity are significantly smaller than the correlation length extracted from the spectral gap ($\xi=19.6,56.5$ for the temperatures under consideration); more importantly, values of $c<2$ in conjunction with large $\sigma$ values (indicating the onset of criticality) imply - from the known properties of the polylogarithm function - a divergent average bubble length, which directly contradicts the exact sum rule (cf. above). \par
The reason behind these difficulties becomes obvious if one looks at the ratios $P_n/P_{n-1}$, shown in the right panel of Fig.\ref{fig:PBD}. It then becomes clear that small bubbles have an entirely different behavior than larger bubbles. If one restricts attention to smaller bubbles, e.g. $n<8$ for $T=0.76$ (a temperature quite close to $T_c$), the {\em apparent} asymptotic value of the ratio (cf. dotted line) appears much smaller; the physics behind this is that it takes a free energy which is typically higher than $T$ in order to generate such a bubble. Note that the low apparent values of the exponent $c$ derived from small bubbles (e.g. $c=0.75$ at $T=0.76$) are irrelevant in this context, because they are not accompanied by a sequence of apparent $\Delta f_b$'s approaching zero at the critical temperature - hence no divergences in the either one of the series (\ref{eq:boundsites}) or (\ref{eq:unboundsites}) are generated. 

A proper analysis of the asymptotics of large bubbles (cf. inset of Fig.\ref{fig:PBD}) shows that the ratios $P_n/P_{n-1}$ lead to $\Delta f_b$'s with the correct limiting behavior, i.e. linearly vanishing at $T_c$ (filled squares in Fig. \ref{fig:Df_PBD}). Moreover, the values of $c$ associated with the true asymptotics are also much larger (cf. inset in Fig. \ref{fig:Df_PBD}). At a temperature very near $T_c$, an estimate $c=2.9$ is obtained, clearly consistent with the finite average bubble size demanded by a first-order transition (cf. above).
 
\section{An alternative procedure}
It is possible to use simple linear algebra in order to extract some formal properties of the limit of very large bubbles $n \to \infty$.
Using the overlap matrix elements
\begin{equation}
	B_{\nu \nu'} =  \int_{-\infty}^{y_c} dy \>  	\phi_{\nu}(y) \phi_{\nu'}(y)
\end{equation}
it is possible to rewrite (\ref{eq:bubbleprob}) in the form
\[
P_n = < A |  C \cdots C | A > 
\]
where the matrix product contains $n$ factors, $C_{\nu \nu'}=(\Lambda_\nu \Lambda_\nu'/\Lambda_0^2)^{1/2} (\delta_{\nu \nu'}- B_{\nu \nu'})$
and $A_\nu=( \Lambda_\nu / \Lambda_0)^{1/2} B_{0\nu} $. It then follows that
\begin{equation}
	P_n = \sum_{\alpha} \mu_\alpha^n |a_\alpha|^2
\end{equation}
where $\{\mu_\alpha \}$ are the eigenvalues of the real symmetric matrix $C$ and $a_\alpha=(SA)_\alpha$, where $S$ 
is the orthogonal matrix which diagonalizes $C$.

If the spectrum of $C$ is continuous  - as the numerical computations suggest - then it is possible to rewrite the bubble probability in a continuum form 
\begin{equation}
	P_n = \mu_0^n \int_0^\infty dx \> {\cal R}(x) \beta (x) e^{- xn/T} 
\end{equation}
where the eigenvalues are now labeled as $\mu_\alpha \to \mu_0 e^{-x/T}$ with a density $ {\cal R}(x)$, and $|a_\alpha|^2 \to \beta(x)$. In the limit 
of very large $n$ only a very narrow range $x<T/n$ contributes to the integral. Therefore, if ${\cal R}(x) \propto x^{-\zeta}$ and $\beta(x) \propto x^\eta$ in the neighborhood of zero, it follows that 
\[
P_n   \propto \left( \frac{T}{n}\right)^{\eta-\zeta+1} \mu_0^n 
\]
and, by comparison with  (\ref{eq:PnPS}), we conclude that in the PBD model the exponent $c=\eta-\zeta+1$ reflects the behavior of the density of eigenvalues of $C$ and the $\beta$ function near the high end of the spectrum. Furthermore, the highest eigenvalue $\mu_0$ can be used to extract $\Delta f_b/T = 1/\sigma = -\ln \mu_0 $. \par
\begin{figure}[h!]
\resizebox{0.482\textwidth}{!}
{\includegraphics{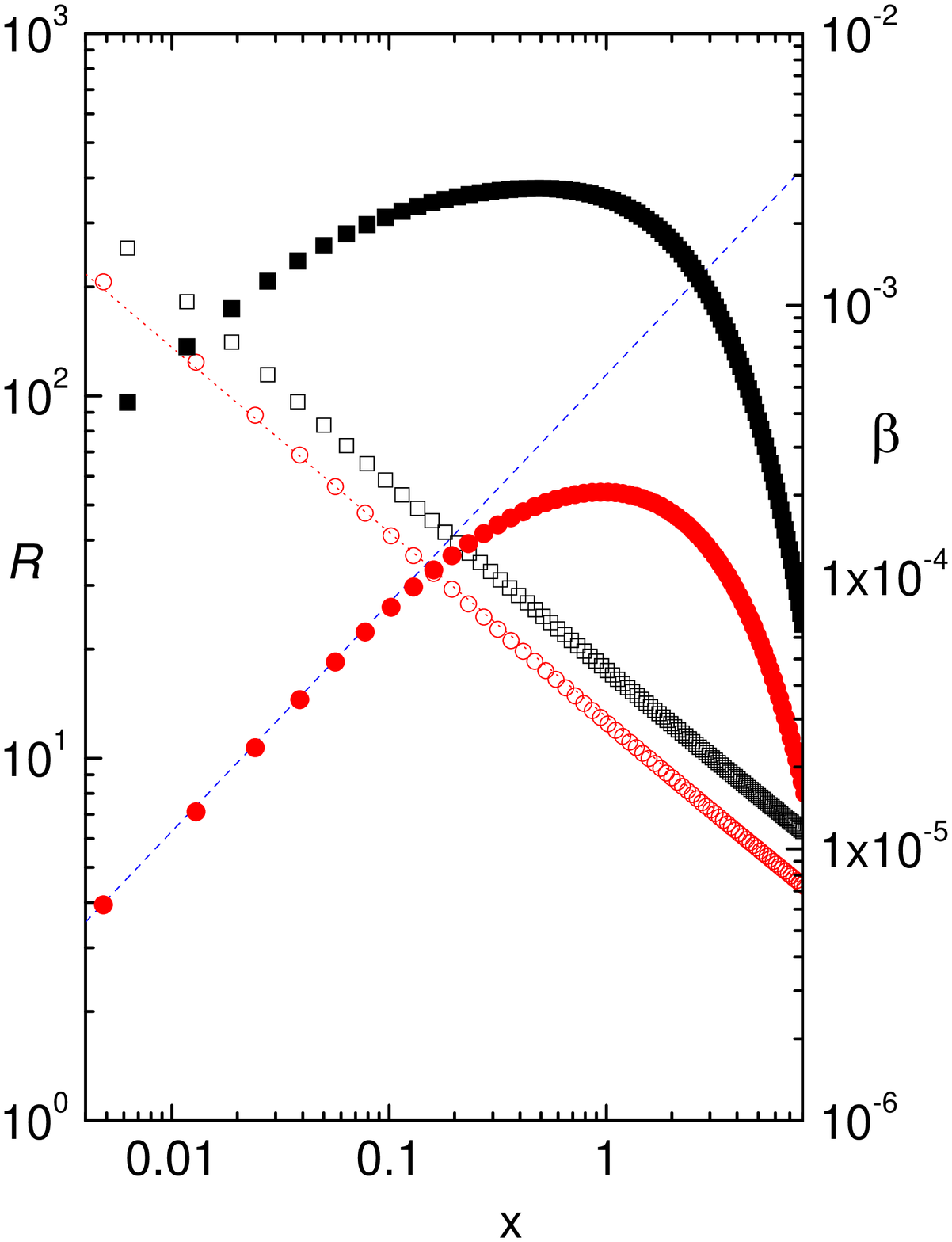}	
\includegraphics{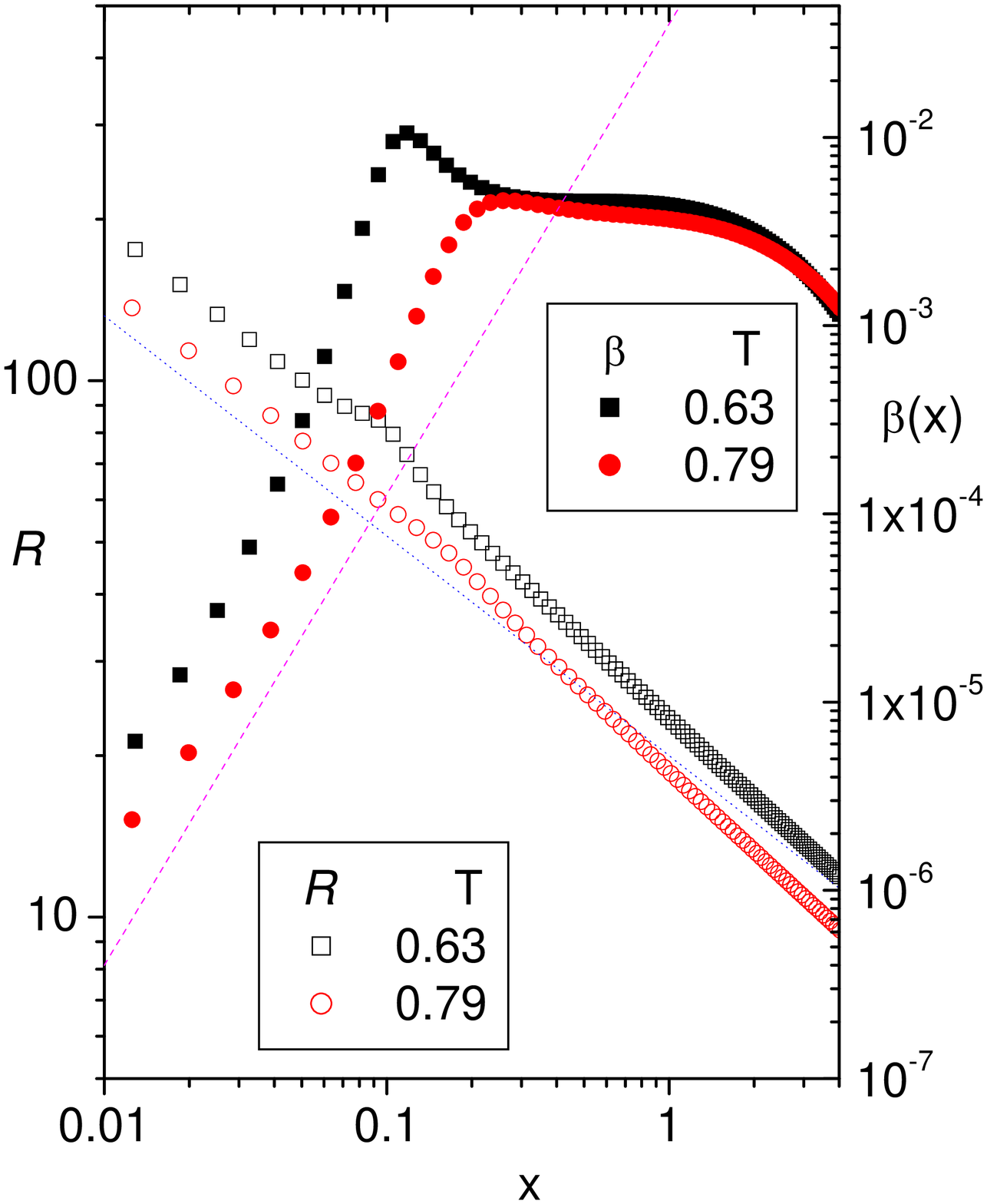}}	
\vskip -.5truecm
\caption{ 
{\small (Color online) The functions ${\cal R}(x)$ (open symbols, left y-axis) and $\beta(x)$ (filled symbols, right y-axis).
{\em Left panel}, $\rho=0$: $T=0.85$ (squares), $T=1.20$ (circles). The density of eigenvalues can be fitted with a power law 
$\zeta=0.51$, regardless of temperature; the function $\beta(x)$ also exhibits a power-law behavior as $x \to 0$; at the highest temperature a value $\eta=0.81$ is obtained.
{\em Right panel}:  $\rho=1$: $T=0.63$ (squares), $T=0.79$ (circles). The density of eigenvalues exhibits a slight anomaly around $x=.1$; in the region $x \to 0$ it follows a power law $\zeta=0.41$ (slope of dotted line, guide to the eye). The function $\beta(x)$ exhibits a strong anomaly around $x=0.1$ and does not seem to settle to a pure power law behavior; however, the {\em effective slope} is quite high (dashed line, guide to the eye has a slope of 2.5).  
}}
\label{fig:DOEetc}
\end{figure}
Fig. \ref{fig:DOEetc} illustrates the behavior of the functions ${\cal R}(x)$ and $\beta(x)$ in 
the cases of both linear and nonlinear stacking. In the first case (left panel), the asymptotics lead to a value $c=1.33$ for $T=1.20$, which is close, but not identical with the $1.47$ obtained by the extrapolation procedure in the previous section. In the nonlinear stacking case (right panel), it is seen that the crossover from small to larger bubbles has its origins in a strong anomaly of the $\beta(x)$ function near $x=0.1$. Using an effective exponent $\eta=2.5$, and $\zeta=0.4$ would imply $c=3.1$, in reasonable agreement with the extrapolation estimates of the previous section.
 
\section{Bubble shapes}
The alternative method described in the previous section, owing to its superior computational efficiency, is uniquely suited to deal with the repeated calculations involved in obtaining full (average) displacement profiles of bubbles. A straightforward generalization of the above scheme allows the calculation of the conditional average displacement of the $s$th site in a bubble of size $n$,
\begin{equation}
	{\bar y}(s|n)  = \frac{1}{P_n} \sum_{\alpha,\alpha'} a_\alpha \mu_\alpha^{s-1} {\tilde D}_{\alpha,\alpha'}
	\mu_{\alpha'}^{n-s}a_{\alpha'}
	\label{eq:ycondav}
\end{equation}
where ${\tilde D}=SDS^{-1}$ and 
\begin{equation}
	D_{\nu \nu'} =  \int_{y_c}^{\infty} dy \>  	\phi_{\nu}(y) \> y \> \phi_{\nu'}(y)  \quad.
\end{equation}
Note that the double sum in (\ref{eq:ycondav}) is - apart from a factor $y_s$ in the integrand - essentially the statistical weight (\ref{eq:bubbleprob}).

\begin{figure}[h!]
\resizebox{0.482\textwidth}{!}
{\includegraphics{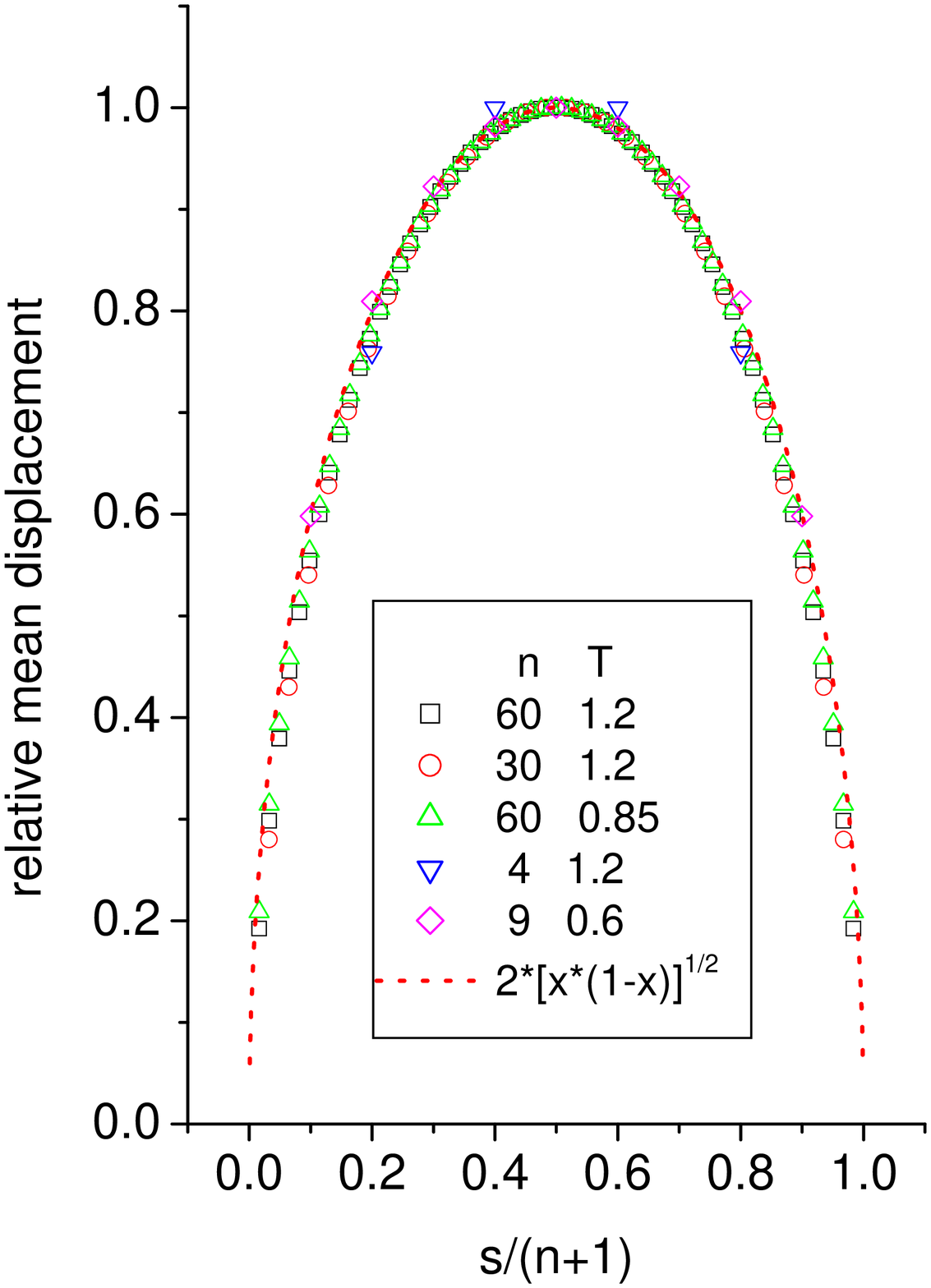}	
\includegraphics{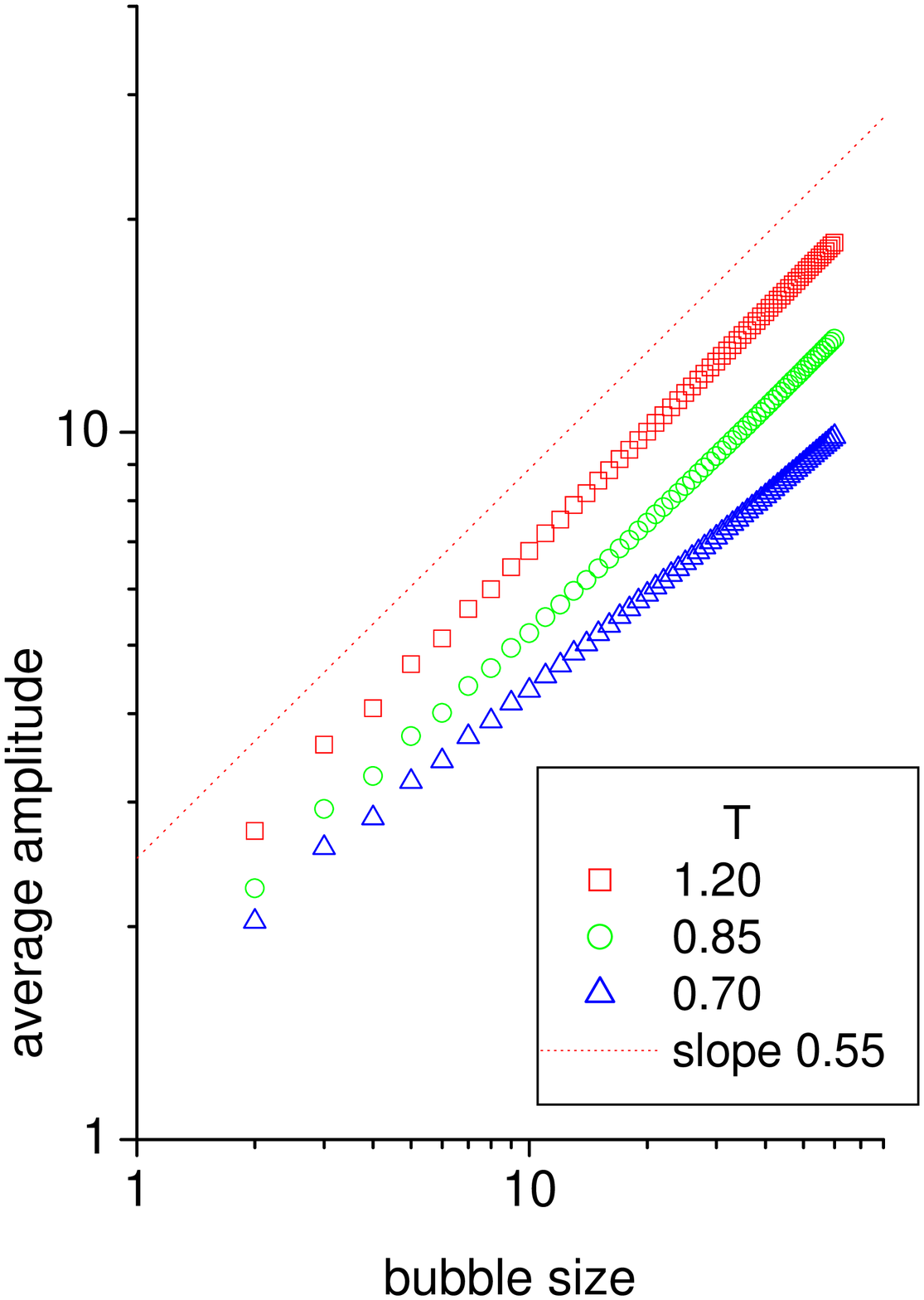}}	
\vskip -.5truecm
\caption{ 
{\small 
(Color online) 
Bubble shapes and sizes in the case $\rho=0$. {\em Left panel:} Relative average displacements of sites in a bubble vs. relative site coordinate. Results from a variety of sizes and temperatures collapse on a single curve. 
  {\em Right panel:} The maximal displacement of an $n-$site bubble for various temperatures.   
}}
\label{fig:BubbShape2nd}
\end{figure}

Fig. \ref{fig:BubbShape2nd} summarizes the results obtained via this approach for bubble shapes in the case of vanishing nonlinear stacking. The left panel shows that if the reduced average displacements (i.e. divided by the maximal displacement found for each bubble) are plotted against the relative site coordinate $s/n$, the shapes obtained are independent of bubble size and/or temperature. Results are well fitted by the shape 
\begin{equation}
{\bar y}(s|n) = y_n^0 (T) [4x(1-x)]^{\tau}  \quad, 
	\label{eq:shape}
\end{equation}
	where $y_n^0(T)$ can be understood as an average amplitude of an $n-$site bubble and $\tau=1/2$. The dependence of bubble amplitudes on size and temperature is shown in the right panel. The amplitudes vary with size according to a power law with an exponent close to $1/2$. The overall temperature dependence appears to be roughly linear.\par

A note of caution is due at this point. The results described in this paper do not demonstrate the existence of a bubble as a well-defined long-lived entity of a given fixed size. The interpretation of the shape reported here is somewhat more indirect, since it concerns statistical average profiles of fluctuating objects. On the other hand {\em if} a time-dependent solution of a finite extent exists then its time-averaged spatial profile, after allowing for corrections due to interactions with phonons, should look like the left panel of Fig. \ref{fig:BubbShape2nd}. 

Long-lived entities with an internal oscillation (discrete breathers (DB)), have been reported \cite{AubryDBrev} in the class of models under consideration here.  Unfortunately, most of the work done on DBs concerns objects which are very localized in space. It would be interesting to examine whether approximate breather-like excitations, perhaps with shorter lifetimes and extending over many lattice sites, could produce average displacement profiles compatible with (\ref{eq:shape}).

\begin{figure}[h]
\resizebox{0.482\textwidth}{!}
{\includegraphics{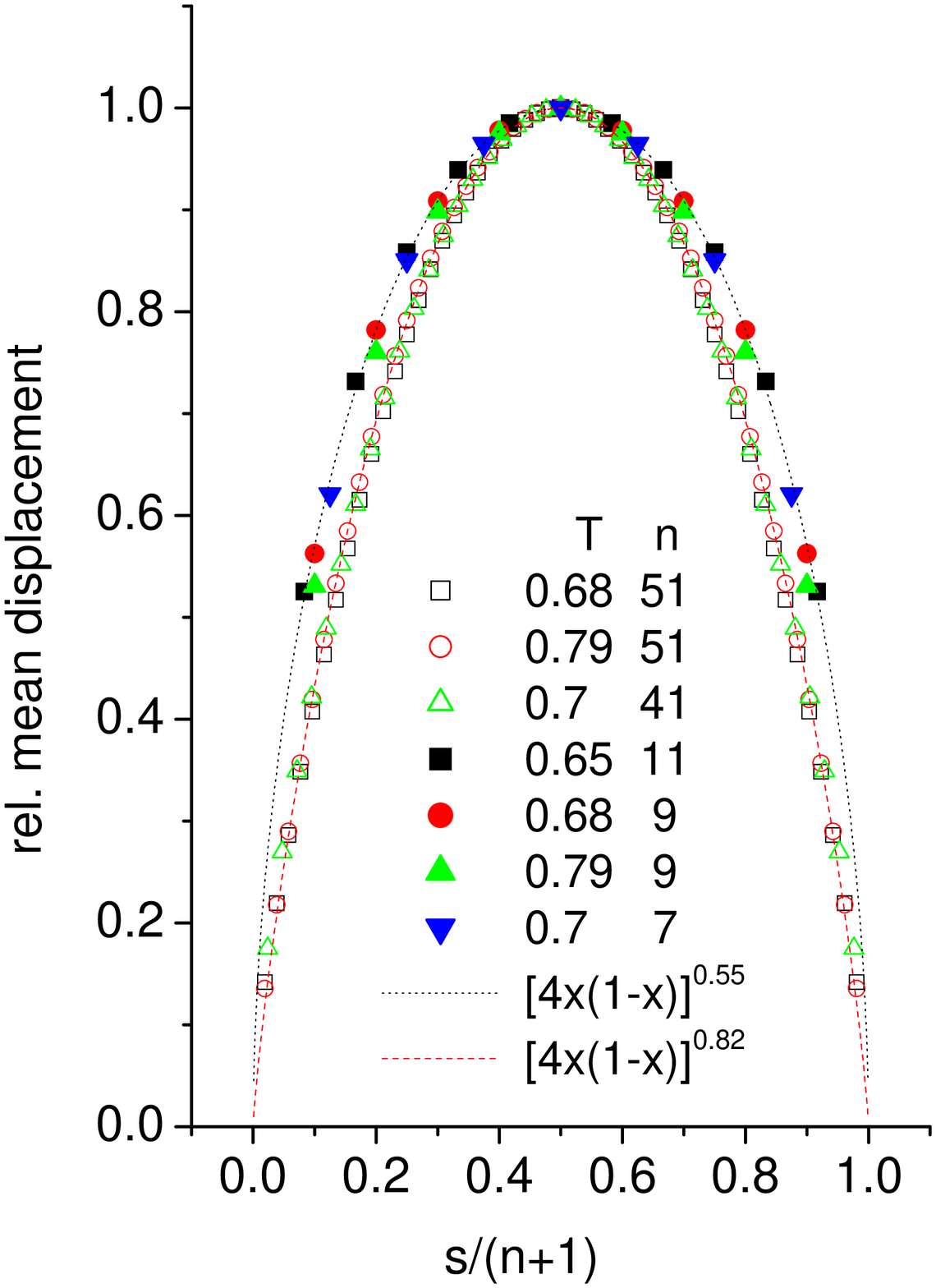}	
\includegraphics{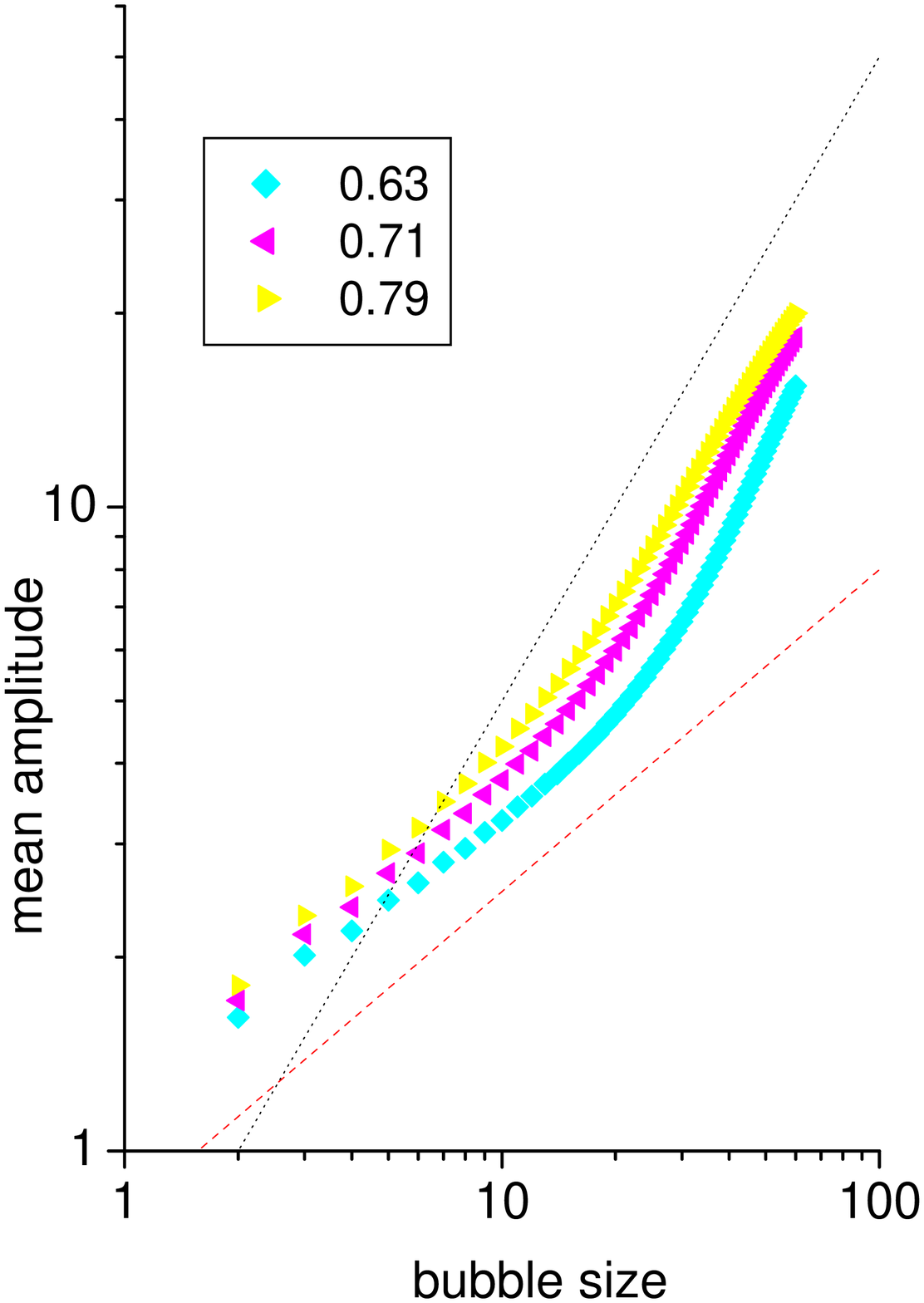}}	
\vskip -.5truecm
\caption{ 
{\small  
(Color online) 
Bubble shapes and sizes in the case $\rho=1$. {\em Left panel:} Relative average displacements of sites in a bubble vs. relative site coordinate. Results from different temperatures and sizes collapse on different curves for large (open symbols) and small (filled symbols) bubbles. 
  {\em Right panel:} The maximal displacement of an $n-$site bubble for various temperatures. Note the crossover which occurs around $n=10-20$ at temperatures both near and far from $T_c$.
}}
\label{fig:BubbShape1st}
\end{figure}
In the case of nonlinear stacking, the results of the previous section indicate that small and large bubbles behave differently. The derived shapes (Fig. \ref{fig:BubbShape1st}, right panel) confirm this dichotomy. Small-size bubbles tend to have a slightly different shape from large-size bubbles. The difference can be expressed in terms of the exponent $\tau$ of the size function (\ref{eq:shape}), with $\tau=0.55$ for small-size and $\tau=0.82$ for large-size bubbles. 
A similar dichotomy occurs when we look at the dependence of bubble amplitude vs. size. There is a crossover behavior from a low-exponent to a high-exponent region (cf. the dotted and dashed lines with slopes $1/2$ and $1$ respectively serving as guides to the eye). This type of behavior complements the findings on the incremental free energy needed for bubble growth.  Amplitude grows weakly with size for small bubbles. After a bubble size threshold ($10-20$ sites) has been crossed  growth becomes a lot easier.  

\section{Summary and discussion}
The detailed properties of locally denatured regions (bubbles) of homogeneous DNA have been discussed in the framework of the PBD model. It has been shown that bubble statistics is very sensitive to the presence of nonlinear stacking interactions. In summary: 

In the limit of linear base stacking ($\rho=0$)
the distribution of bubble sizes can be described by the product of an exponential and a power law (cf. Eq. \ref{eq:PnPS}). The latter is characterized by an exponent $c$ which is weakly dependent on temperature. The exponent's value near the critical temperature, close to $3/2$,  suggests a formal analogy with the Poland-Scheraga description of $3$-$d$ polymer loops in the random walk limit. However, the analogy does not rest on microscopic footing; the logarithmic correction to the bubble entropy is not related to looping and, more importantly, it is not even approximately constant in temperature; on the other hand, the value of $c$ {\em must} approach $3/2$ at the critical temperature since the transition is known to be exactly second order in the case $\rho=0$.
\begin{figure}[h]
\centering
	\vskip -.3truecm
\resizebox{0.35\textwidth}{!}
{		\includegraphics{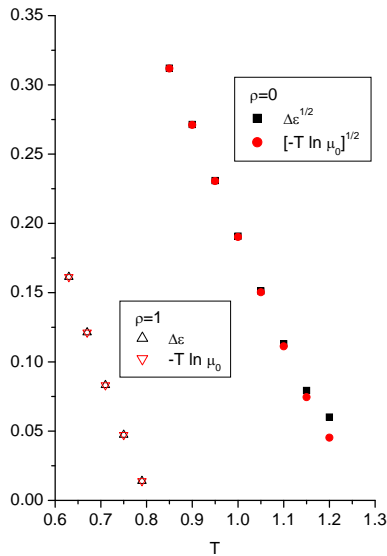}}
\caption{ {\small 
(Color online) 
Comparison of $\Delta f_b=-T\ln \mu_0$ obtained in the context of the associated eigenvalue problem with the TI spectral gap
$\Delta \epsilon$ in the cases $\rho=1$ (open symbols) $\rho=0$ (filled symbols); in the latter case the square root is plotted.
}}
 	\label{fig:EigComp}
\end{figure}

In the presence of nonlinear base stacking ($\rho=1$) the analysis of successive probability ratios $P_n/P_{n-1}$ reveals a more complex behavior of the exponent $c$. Depending on the range of bubble sizes analyzed, different {\em apparent} values of $c$ are estimated. Small bubbles lead to small values of $c$, larger bubbles suggest larger $c$ values; the latter are of course consistent with the apparent first-order transition. The threshold seems also to control average bubble shapes. Note that it is possible to relate the value of the threshold to the parameters of the thermal barrier $U(y) = (T/2) \ln(1+ \rho e^{-2by})$ \cite{TDP,Hwa} induced by the nonlinear base stacking interaction. For $\rho=1$ the barrier becomes effective at displacements of order $1/(2b) \sim 7$. This corresponds to the region where the crossover in slope occurs in the right panel of Fig. \ref{fig:BubbShape1st} and explains the relative insensitivity of the bubble size threshold to temperature.

A formal point deserves to be mentioned. The values of $\Delta f_b$ obtained in the framework of the associated eigenvalue problem (Section IV), appear to be identical with those of the spectral gap $\Delta \epsilon$ (cf. Fig. \ref{fig:EigComp}). Differences, in the case of $\rho=0$, are of the order of the inverse matrix size used in the TI calculation and are more pronounced where the expected critical rounding of eigenvalues occurs; for $\rho=1$ - where there is no observable critical rounding of eigenvalues - the difference vanishes. The property $\Delta f_b = \Delta \epsilon$ should hold in the limit of infinite matrix size; this can be further confirmed by perturbational estimates of the spectrum of the $C$ matrix. Bubble statistics thus offers a physical interpretation of the TI spectral gap as the limiting $(n \to \infty)$ free energy which must be provided in order to achieve an incremental growth of a bubble by a single site.

I would like to acknowledge the detailed and constructive comments made by one of the anonymous referees and thank George Kalosakas for sending me a copy of Ref. \cite{Kalo2007}.


\begin{thebibliography}{0}
\bibitem{foot0} 
The astute reader will note that the probability of small bubble formation is higher at lower temperatures. This apparently counterintuitive property is due to the definition of a bubble, which demands that the two ends are bound. $P_n$ is a probability of a configuration of an $n+2$ site group; it reflects the competition between holding the ends bound (higher probability at lower temperatures) and softening the bonds of the $n$ interior sites (higher probability at higher temperatures).
\bibitem{foot2}
The form (\ref{eq:PnPS}) has been derived in the context of the Poland-Scheraga\cite{PoSche} model, with the exponent $c$ associated with the entropy of loops (bubbles) modeled as self-avoiding polymer chains. There is no a priori reason why the result of such a geometrically based description - which mainly reflects the dimensionality of space - should correspond to any feature of the one-dimensional nonlinear lattice dynamics of the PBD model (cf. however below for some general arguments and a more detailed description of the meaning of the exponent $c$ in the PBD context).
\end{thebibliography}
\end{document}